\documentclass{article}
\usepackage{spconf,amsmath,graphicx,hyperref}
\usepackage{amssymb}   
\usepackage{setspace}
\usepackage{booktabs}
\usepackage{multirow}

\title{Domain-Adaptive Model Merging across Disconnected Modes}

\name{Junming Liu$^{1}$, Yusen Zhang$^{1}$, Rongchao Zhang$^{2}$, 
      Wenkai Zhu$^{3}$, Tian Wu$^{4}$\sthanks{Corresponding author.}}

\address{$^{1}$Tongji University \quad
         $^{2}$Peking University \quad
         $^{3}$Southeast University \quad
         $^{4}$Nanchang University \\
         \small \texttt{liu\_junming6917@tongji.edu.cn, wutian@ncu.edu.cn}}

\begin{document}
%
\maketitle
\begin{abstract}
Learning across domains is challenging when data cannot be centralized due to privacy or heterogeneity, which limits the ability to train a single comprehensive model. Model merging provides an appealing alternative by consolidating knowledge from multiple specialized models into one, avoiding data sharing and reducing retraining cost. In this work, we present DMM, a data-free model merging framework designed to handle highly divergent models. DMM proceeds in three steps. First, domain-specific models are trained independently. Second, models with high similarity are merged using standard techniques to ensure stability. Third, we synthesize pseudo-data from normalization statistics and distill knowledge from divergent models into the merged model through a lightweight refinement guided by these samples.
This approach preserves rare but critical knowledge while maintaining stability. Extensive experiments on unimodal and multimodal benchmarks show that DMM achieves state-of-the-art performance over existing merging methods.
\end{abstract}
\section{Introduction}
\label{sec:intro}

The rapid expansion of machine learning applications across diverse domains has led to a growing demand for methods that can efficiently adapt knowledge without relying on centralized training \cite{Sarker_2021_ML, Yang_2024_Model}. In many practical scenarios, data remains fragmented due to privacy regulations \cite{Shokri_2015_Privacy}, acquisition costs \cite{Li_2021_Data}, or domain heterogeneity \cite{Liu_2020_Hete}, making it difficult to train a single model on all available data \cite{Mcmahan_2017_FedAvg, Liu_2025_FedRecon, Liu_2025_Mosaic}.
As a result, combining multiple specialized domain-specific models into a unified one has become an attractive strategy for transferring and consolidating knowledge \cite{Yadav_2023_TIES}.

Model merging offers a promising solution in this context \cite{Ainsworth_2023_Git, Xu_2024_Training}. Instead of retraining from scratch, it leverages existing models that are already optimized for specific domains or tasks and integrates them into a single model with enhanced generalization ability. This paradigm reduces computational cost, circumvents the need for data sharing, and provides a scalable approach for adapting models across domains \cite{Yang_202_Representation}.

Despite its advantages, model merging faces several critical challenges. Many approaches assign merging weights according to the size of the training data \cite{Mcmahan_2017_FedAvg}. Such strategies risk suppressing models trained on scarce yet valuable samples, causing the merged model to overlook rare but highly discriminative patterns. Other approaches rely on parameter similarity \cite{Nasery_2025_PLeaS, Sun_2025_Cat}, assuming that models lie within the same optimization basin \cite{Ainsworth_2023_Git}. While effective for similar models, this assumption fails when models are highly divergent. In these cases, dissimilar models are often down-weighted or excluded altogether to preserve stability, which discards potentially important domain-specific knowledge. Furthermore, some merging methods still require auxiliary data or retraining to achieve convergence \cite{Wei_2025_Representation}, limiting their applicability in truly data-free or resource-constrained environments.

To address these issues, we propose \textbf{DMM}, a data-free model merging framework that explicitly accounts for divergent models while maintaining stability. DMM operates in three stages. First, domain-specific models are trained independently. Second, models with high similarity are merged using standard techniques, ensuring stable consolidation. Third, instead of discarding the most divergent models, DMM reintegrates them by synthesizing pseudo-data from the normalization statistics of these models, and transferring their unique knowledge to the merged model through a lightweight distillation stage \cite{Zhu_2021_DFKD}. This procedure requires only a few steps of fine-tuning without any access to the original training data, while a novel buffer-level update further corrects statistical mismatches across models. This design enables DMM to capture both stable and rare knowledge, improving performance on challenging domains without violating data-free constraints.
We validate DMM on both unimodal and multimodal benchmarks, including image classification and image-text tasks. Results demonstrate that our method consistently outperforms existing merging techniques and achieves new state-of-the-art performance, particularly in settings with imbalanced data distributions.

Our main contributions are summarized as follows. 
First, we propose a buffer-level merging method and provide theoretical guarantees of its effectiveness in capturing global statistics. 
Second, we introduce a lightweight strategy that synthesizes pseudo-data from normalization statistics to distill knowledge from divergent models, enabling the merged model to retain rare but critical information in a fully data-free manner. 
Third, we evaluate DMM on unimodal and multimodal benchmarks, where it consistently outperforms prior methods and achieves state-of-the-art performance.

\section{Preliminary}
\label{sec:pre}

\textbf{Notations.} 
Let $W_0 = \{ W^l_0 \}_{l=1}^L$ denote the parameters of a pretrained network with $L$ layers, where $W^l_0$ represents the parameters of the $l$-th layer. The model is fine-tuned independently on $K$ domains with datasets $\mathcal{D}_1, \dots, \mathcal{D}_K$, yielding domain-specific parameters $W_1, \dots, W_K$. For domain $k$, the parameter offset is defined as the difference between the fine-tuned parameters and the pretrained initialization:
\begin{equation}
\Delta W_k = W_k - W_0 = \{ W^l_k - W^l_0 \}_{l=1}^L.
\end{equation}
The dimensions of $W^l_0$ and $\Delta W^l_k$ depend on the layer type: for a linear layer, $W^l_0, \Delta W^l_k \in \mathbb{R}^{d_l \times d_{l+1}}$, while for elementwise parameters (e.g., normalization scales or biases), $W^l_0, \Delta W^l_k \in \mathbb{R}^{d_l}$.

\textbf{Model merging.} 
The goal is to obtain a merged model $M$ that integrates knowledge from all $K$ fine-tuned models. A basic strategy, often referred to as \emph{parameter arithmetic} \cite{Sun_2025_Cat}, directly adds parameter offsets to the pretrained parameters:
\begin{equation}
M = W_0 + \sum_{k=1}^K \alpha_k \Delta W_k,
\end{equation}
where $\alpha_k$ are scalar coefficients controlling the contribution of each model. This linear combination provides a simple yet effective foundation for model merging, though it may suffer from conflicts when offsets interfere.

\section{Methods}
\label{sec:method}

\begin{figure*}[t]
    \centering
    \fbox{%
        \includegraphics[width=\textwidth]{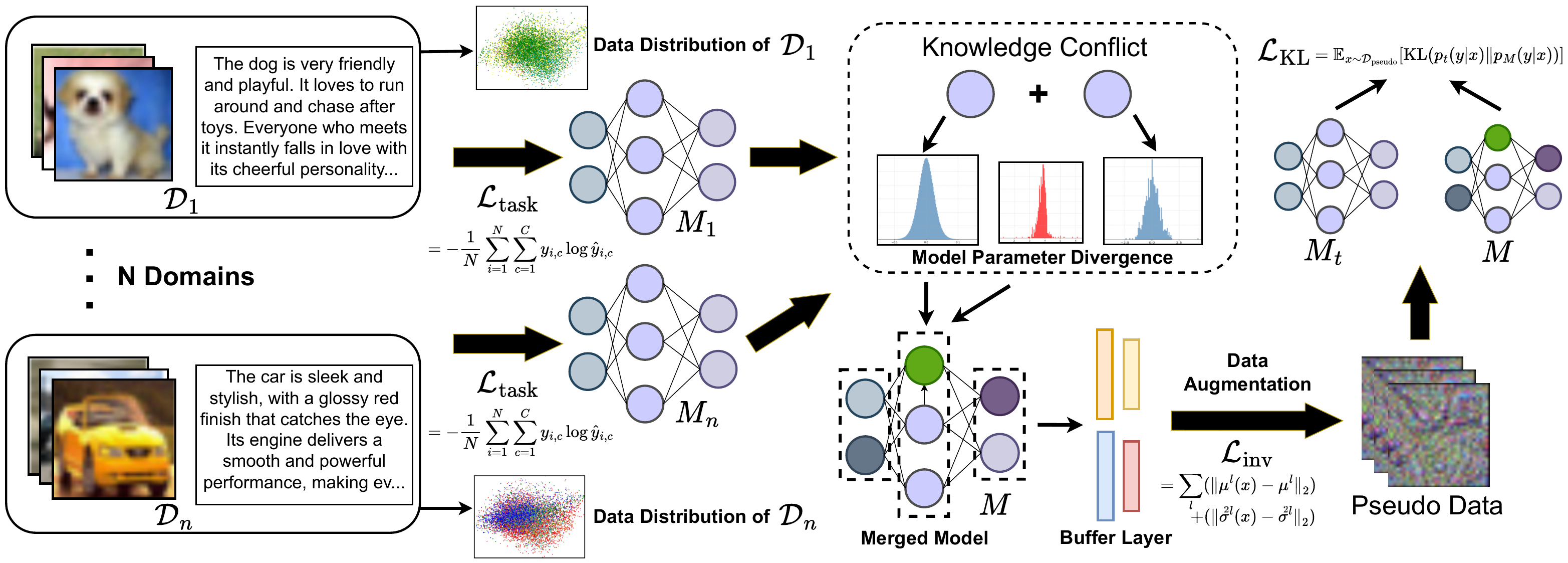}%
    }
    \caption{
        Overall workflow of the proposed DMM framework, 
        which integrates model initialization, model merging with proxy data synthesis, 
        and lightweight data-free knowledge distillation.
    }
    \label{fig:workflow}
\end{figure*}

Our method consists of three components. First, we train unimodal and multimodal models on their respective tasks to obtain well-initialized models. Second, we perform model merging using both parameter aggregation and buffer-level statistics alignment, followed by normalization inversion to synthesize proxy data. Finally, to mitigate knowledge conflicts \cite{Sun_2025_Cat} introduced by merging, we design a lightweight data-free knowledge distillation scheme that leverages high-confidence predictions from divergent models to guide the merged model. Together, these steps enable effective integration of multiple models without access to original training data. The overall workflow is illustrated in Figure~\ref{fig:workflow}.

\subsection{Model Training}

We first describe the training procedure for domain-specific models. 
For unimodal vision tasks, we adopt standard convolutional architectures such as ResNet \cite{He_2016_ResNet} and optimize them using cross-entropy loss for image classification. 
For multimodal classification, we utilize ResNet for image feature extraction and BERT \cite{Devlin_2019_BERT} for text representation, followed by a linear classification layer that maps the concatenated features into the label space. 
Each model is trained independently on its respective dataset, resulting in domain-specialized networks that capture complementary knowledge. 

To consolidate these models, we consider general parameter-averaging strategies. 
In particular, simple approaches such as parameter averaging in federated learning  \cite{Mcmahan_2017_FedAvg, Liu_2025_FedRecon, Liu_2025_Mosaic} or task arithmetic \cite{Xu_2024_Training, Nasery_2025_PLeaS, Sun_2025_Cat} provide a baseline. 
However, naive averaging often leads to conflicts due to domain shifts, motivating our subsequent designs.

\subsection{Buffer Aggregation and Data Inversion}

Modern neural networks equipped with normalization layers maintain running statistics such as mean and variance during training. These buffers encapsulate domain-specific characteristics and can serve as proxies for the underlying data distribution.

Given $K$ models with BN layers indexed by $l=1,\dots,L$, let their buffers be
$\Gamma_k^l = \{\mu_k^l, \sigma_k^l, n_k^l\}$,
where $\mu_k^l$, $\sigma_k^l$, and $n_k^l$ are the mean, standard deviation, and number of tracked batches for model $k$.
The merged statistics are computed as:
\begin{equation}
N^l = \sum_{k=1}^K n_k^l, \quad
\mu^l = \frac{1}{N^l} \sum_{k=1}^K n_k^l \mu_k^l,
\end{equation}
\begin{equation}
(\sigma^l)^2 =
\frac{1}{N^l} \sum_{k=1}^K
\Big(
n_k^l (\sigma_k^l)^2
+
n_k^l (\mu^l - \mu_k^l)^2
\Big).
\end{equation}

Once aggregated, these statistics allow us to reconstruct pseudo-data via distributional inversion. Inspired by DeepInversion \cite{Yin_2020_DeepInversion}, we optimize an input $x$ such that its batch-normalized activations match the global statistics:
\begin{equation}
\mathcal{L}_{\text{inv}}(x) = \sum_{l} \Big( \| \mu^l(x) - \mu^l \|_2 + \| (\sigma^l(x))^2 - (\sigma^l)^2 \|_2 \Big),
\end{equation}
where $\mu^l(x)$ and $(\sigma^l(x))^2$ are the empirical mean and variance of activations at layer $l$ for the synthetic input $x$. This produces lightweight pseudo-data that reflect the global distribution, without access to the original training samples.

\subsection{Data-Free Knowledge Distillation for Conflict Resolution}
While parameter merging provides a starting point, conflicts inevitably arise when models encode divergent knowledge across domains~\cite{Sun_2025_Cat}. 
A common strategy is to down-weight parameters from conflicting models \cite{Nasery_2025_PLeaS}, which improves stability but may discard rare yet valuable knowledge. 
Instead, our goal is to selectively transfer such information into the merged model.

We first compute a divergence score $\tau_k$ for each model $W_k$ with respect to the merged model $M$, combining (i) the averaged neuron-wise dissimilarity in parameter space and (ii) domain heterogeneity inferred from buffer statistics \cite{Ainsworth_2023_Git}. Models with $\tau_k$ exceeding a threshold $\tau$ are treated as extreme outliers, capturing unique but potentially unstable knowledge.

To exploit this knowledge, we employ a data-free knowledge distillation step using the pseudo-data from the inversion stage. Specifically, for a divergent teacher model $M_t$ and the merged student $M$, we minimize:
\begin{equation}
\mathcal{L}_{\text{KD}} = \mathbb{E}_{x \sim \mathcal{D}_{\text{pseudo}}} 
\big[ \text{KL}\big( p_{M_t}(y|x) \; \| \; p_M(y|x) \big) \big],
\end{equation}
where $p_{M_t}(y|x)$ and $p_M(y|x)$ denote the softmax outputs of the teacher and student, respectively. To ensure reliability, we only retain samples where the teacher exhibits high confidence (i.e., $\max_y p_{M_t}(y|x)$ is large) while the student remains uncertain (i.e., high entropy). These samples highlight domain-specific knowledge that would otherwise be lost.

This refinement is highly efficient: pseudo-data generation is lightweight, the distillation requires only a few steps, and no real data or expensive generative models (e.g., GANs or diffusion models) are involved. As a result, our framework preserves privacy, minimizes computational cost, and effectively consolidates both common and rare domain knowledge.

\begin{table*}[t]
\centering
\caption{Accuracy (\%) comparison on CIFAR-10, CIFAR-100, and CrisisMMD under different Dirichlet distributions.}
\label{tab:main_results}
\begin{tabular}{lccccccccc}
\toprule
\multirow{2}{*}{Method} 
& \multicolumn{3}{c}{CIFAR-10} 
& \multicolumn{3}{c}{CIFAR-100} 
& \multicolumn{3}{c}{CrisisMMD} \\
\cmidrule(lr){2-4} \cmidrule(lr){5-7} \cmidrule(lr){8-10}
& $\alpha=1.0$ & $\alpha=0.1$ & $\alpha=0.01$ 
& $\alpha=1.0$ & $\alpha=0.1$ & $\alpha=0.01$ 
& $\alpha=1.0$ & $\alpha=0.1$ & $\alpha=0.01$ \\
\midrule
FedAvg \cite{Mcmahan_2017_FedAvg}       & 78.34 & 56.32 & 36.76 & 65.55 & 59.30 & 48.72 & 60.93 & 45.21 & 22.50 \\
FedProx \cite{Li_2020_FedProx}          & 79.45 & 56.73 & 37.32 & 65.54 & 59.64 & 49.73 & 62.88 & 46.84 & 21.32 \\
FedBN \cite{Li_2021_FedBN}              & 79.52 & 57.68 & 38.35 & 66.92 & 59.16 & 50.70 & 60.51 & 44.79 & 23.72 \\
    FedAvg+DMM                              & \textbf{82.30} & \textbf{65.85} & \textbf{53.66} & \textbf{68.68} & \textbf{64.31} & \textbf{53.04} & \textbf{64.33} & \textbf{49.70} & \textbf{30.46} \\
\midrule
PLeaS \cite{Nasery_2025_PLeaS}          & 60.51 & 45.77 & 33.30 & 49.74 & 44.97 & 39.56 & 52.18 & 40.11 & 23.67 \\
Git Re-Basin \cite{Ainsworth_2023_Git}  & 55.94 & 43.10 & 31.96 & 46.68 & 44.80 & 41.68 & 50.34 & 42.29 & 24.76 \\
Cat-Merge \cite{Sun_2025_Cat}           & 59.22 & 47.67 & 36.89 & 50.18 & 45.51 & 41.11 & 55.03 & 43.55 & 25.29 \\
Cat-Merge+DMM                           & \textbf{65.01} & \textbf{55.86} & \textbf{48.63} & \textbf{53.00} & \textbf{48.42} & \textbf{45.34} & \textbf{57.73} & \textbf{45.45} & \textbf{28.42} \\
\bottomrule
\end{tabular}
\end{table*}

\section{Experiments}
\label{sec:exp}

\subsection{Experimental Setup}

\subsubsection{Datasets}
We evaluate our approach on three benchmarks. 
\textbf{CIFAR-10} and \textbf{CIFAR-100} \cite{Krizhevsky_2009_CIFAR} are standard image classification datasets containing 10 and 100 classes, respectively, with 50,000 training and 10,000 test images each. 
\textbf{CrisisMMD} \cite{Alam_2018_CrisisMMD} is a multimodal dataset consisting of images and accompanying textual reports collected from 18 real-world crisis events. It contains a total of 18,036 image–text pairs annotated with humanitarian categories, enabling evaluation of cross-modal classification tasks.

\subsubsection{Baselines}
We compare our method against representative federated learning and model merging approaches, including: 
\textbf{FedAvg} \cite{Mcmahan_2017_FedAvg}, 
\textbf{FedProx} \cite{Li_2020_FedProx}, 
\textbf{FedBN} \cite{Li_2021_FedBN}, 
\textbf{Cat-Merge} \cite{Sun_2025_Cat}, 
\textbf{PLeaS} \cite{Nasery_2025_PLeaS}, and 
\textbf{Git Re-Basin} \cite{Ainsworth_2023_Git}. 

\subsubsection{Implementation Details}
Each domain's local data is partitioned into $N=10$ subsets using a Dirichlet distribution, controlling the Non-IID degree via the concentration parameter.
For unimodal visual models, we use ResNet-18 as the backbone. 
For multimodal classification, ResNet-18 extracts image features, BERT extracts text features, and a linear layer is used as the final classifier. 
All models are trained with a learning rate of 0.005, batch size of 128, and 20--50 local epochs depending on the dataset, using cross-entropy loss.
Experiments are conducted on a single NVIDIA A100 GPU with 80GB memory using three seeds.

\subsection{Results}
We report the results of our approach in Table~\ref{tab:main_results}. 
Our analysis proceeds along two perspectives: federated aggregation and model merging.

First, for federated learning methods, we observe that integrating DMM into FedAvg leads to substantial improvements, especially under highly Non-IID settings (e.g., $\alpha=0.01$). 
This is because FedAvg suffers significantly from severe data heterogeneity: the aggregated model is poorly aligned, and local models remain highly divergent. 
By leveraging synthetic proxy data for post-aggregation fine-tuning, DMM acts as both a data augmentation strategy and a mechanism for reconciling divergent knowledge across clients.

Second, for model merging baselines, we also see consistent gains after incorporating DMM. 
Although the relative improvements are smaller compared to FedAvg, this is expected: these baselines already account for data heterogeneity in their design. 
Nevertheless, DMM still brings notable boosts, confirming its compatibility with strong merging techniques.

Finally, we observe that as $\alpha$ decreases, data heterogeneity intensifies and overall performance drops, while the gains of DMM become more pronounced. This confirms that DMM is especially effective under high heterogeneity, complementing both aggregation- and merging-based paradigms.

\subsection{Ablation Study}

\begin{table}[t]
\centering
\caption{Accuracy (\%) comparison under moderately Non-IID ($\alpha=0.1$) on CIFAR-10 and CIFAR-100.}
\label{tab:ablation_cifar_01}
\begin{tabular}{lcc}
\toprule
Method & CIFAR-10 & CIFAR-100 \\
\midrule
Cat-Merge (baseline)      & 47.67 & 45.51 \\
+ Buffer Aggregation   & 48.11    & 45.63    \\
+ Inversion Augmentation & 53.32  & 47.16    \\
+ DMM (full)           & 55.86    & 48.42     \\
\bottomrule
\end{tabular}
\end{table}

We conduct ablation studies based on FedAvg and Cat-Merge under both moderately and highly Non-IID settings ($\alpha=0.1$ and $\alpha=0.01$) on CIFAR-10 and CIFAR-100.
Specifically, we evaluate the contributions of three components of our DMM framework: 
(1) buffer-level aggregation strategies, 
(2) data augmentation on inversion-synthesized proxy data (without knowledge distillation), and 
(3) training with synthesized data combined with knowledge distillation. 
On both CIFAR-10 and CIFAR-100, the superiority of the full DMM method is clearly observed, highlighting the effectiveness of integrating all three components.
We further evaluate DMM under varying numbers of domains and observe consistently robust performance as $N$ increases.
The training cost remains comparable to baselines such as FedAvg and Cat-Merge, indicating negligible additional overhead.

\begin{table}[t]
\centering
\caption{Accuracy (\%) comparison under high Non-IID ($\alpha=0.01$) on CIFAR-10 and CIFAR-100.}
\label{tab:ablation_cifar_001}
\begin{tabular}{lcc}
\toprule
Method & CIFAR-10 & CIFAR-100 \\
\midrule
FedAvg (baseline)      & 36.76 & 48.72 \\
+ Buffer Aggregation   & 38.98    & 49.43    \\
+ Inversion Augmentation & 50.38  & 51.10    \\
+ DMM (full)           & 53.66    & 53.04    \\
\bottomrule
\end{tabular}
\end{table}

\section{Conclusion}
In this work, we introduced DMM, a data-free model merging framework tailored for scenarios with strong domain heterogeneity. 
By combining buffer-guided pseudo-data generation with selective knowledge distillation from divergent models, DMM effectively reconciles both common and rare domain-specific knowledge while preserving stability. 
Our experiments across unimodal and multimodal benchmarks demonstrate consistent improvements over state-of-the-art aggregation and merging methods. 
These results highlight the promise of data-free refinement as a practical solution for building robust and unified models in privacy-sensitive and highly heterogeneous environments.

\section{Acknowledgments}

This work was supported in part by Jiangxi Provincial Natural Science Foundation under No. 20252BAC200613; in part by Jiangxi Provincial Early-Career Youth Science and Technology Talent Cultivation Project under No.20244BCE52007.

{\fontsize{9.5pt}{10.5pt}\selectfont
\bibliographystyle{IEEEbib}
\bibliography{strings,refs}

@inproceedings{Mcmahan_2017_FedAvg,
  title={Communication-efficient learning of deep networks from decentralized data},
  author={McMahan, Brendan and Moore, Eider and Ramage, Daniel and Hampson, Seth and y Arcas, Blaise Aguera},
  booktitle={Artificial intelligence and statistics},
  pages={1273--1282},
  year={2017},
  organization={PMLR}
}

@inproceedings{Yadav_2023_TIES,
 author = {Yadav, Prateek and Tam, Derek and Choshen, Leshem and Raffel, Colin A and Bansal, Mohit},
 booktitle = {Advances in Neural Information Processing Systems},
 editor = {A. Oh and T. Naumann and A. Globerson and K. Saenko and M. Hardt and S. Levine},
 pages = {7093--7115},
 publisher = {Curran Associates, Inc.},
 title = {TIES-Merging: Resolving Interference When Merging Models},
 url = {https://proceedings.neurips.cc/paper_files/paper/2023/file/1644c9af28ab7916874f6fd6228a9bcf-Paper-Conference.pdf},
 volume = {36},
 year = {2023}
}

@article{Liu_2025_FedRecon,
  title={Fedrecon: Missing modality reconstruction in distributed heterogeneous environments},
  author={Liu, Junming and Zeng, Guosun and Wang, Ding and Gao, Yanting and Jin, Yufei},
  journal={arXiv preprint arXiv:2504.09941},
  year={2025}
}

@article{Liu_2025_Mosaic,
  title={Mosaic: Data-Free Knowledge Distillation via Mixture-of-Experts for Heterogeneous Distributed Environments},
  author={Liu, Junming and Gao, Yanting and Meng, Siyuan and Sun, Yifei and Wu, Aoqi and Jin, Yufei and Chen, Yirong and Wang, Ding and Zeng, Guosun},
  journal={arXiv preprint arXiv:2505.19699},
  year={2025}
}

@inproceedings{Shokri_2015_Privacy,
author = {Shokri, Reza and Shmatikov, Vitaly},
title = {Privacy-Preserving Deep Learning},
year = {2015},
isbn = {9781450338325},
publisher = {Association for Computing Machinery},
address = {New York, NY, USA},
url = {https://doi.org/10.1145/2810103.2813687},
doi = {10.1145/2810103.2813687},
booktitle = {Proceedings of the 22nd ACM SIGSAC Conference on Computer and Communications Security},
pages = {1310–1321},
numpages = {12},
location = {Denver, Colorado, USA},
series = {CCS '15}
}

@article{Li_2021_Data,
author = {Li, Yifan and Yu, Xiaohui and Koudas, Nick},
title = {Data acquisition for improving machine learning models},
year = {2021},
issue_date = {June 2021},
publisher = {VLDB Endowment},
volume = {14},
number = {10},
issn = {2150-8097},
url = {https://doi.org/10.14778/3467861.3467872},
doi = {10.14778/3467861.3467872},
journal = {Proc. VLDB Endow.},
month = jun,
pages = {1832–1844},
numpages = {13}
}

@InProceedings{Xu_2024_Training,
    author    = {Xu, Zhengqi and Yuan, Ke and Wang, Huiqiong and Wang, Yong and Song, Mingli and Song, Jie},
    title     = {Training-Free Pretrained Model Merging},
    booktitle = {Proceedings of the IEEE/CVF Conference on Computer Vision and Pattern Recognition (CVPR)},
    month     = {June},
    year      = {2024},
    pages     = {5915-5925}
}

@inproceedings{Ainsworth_2023_Git,
title={Git Re-Basin: Merging Models modulo Permutation Symmetries},
author={Samuel Ainsworth and Jonathan Hayase and Siddhartha Srinivasa},
booktitle={The Eleventh International Conference on Learning Representations },
year={2023},
url={https://openreview.net/forum?id=CQsmMYmlP5T}
}

@inproceedings{Yang_202_Representation,
title={Representation Surgery for Multi-Task Model Merging},
author={Enneng Yang and Li Shen and Zhenyi Wang and Guibing Guo and Xiaojun Chen and Xingwei Wang and Dacheng Tao},
booktitle={Forty-first International Conference on Machine Learning},
year={2024},
url={https://openreview.net/forum?id=Sbl2keQEML}
}

@InProceedings{Nasery_2025_PLeaS,
    author    = {Nasery, Anshul and Hayase, Jonathan and Koh, Pang Wei and Oh, Sewoong},
    title     = {PLeaS - Merging Models with Permutations and Least Squares},
    booktitle = {Proceedings of the IEEE/CVF Conference on Computer Vision and Pattern Recognition (CVPR)},
    month     = {June},
    year      = {2025},
    pages     = {30493-30502}
}

@inproceedings{Sun_2025_Cat,
title={{CAT} Merging: A Training-Free Approach for Resolving Conflicts in Model Merging},
author={Wenju Sun and Qingyong Li and Yangliao Geng and Boyang Li},
booktitle={Forty-second International Conference on Machine Learning},
year={2025},
url={https://openreview.net/forum?id=zy7Jw91tdh}
}

@inproceedings{Wei_2025_Representation,
title={Representation Surgery in Model Merging with Probabilistic Modeling},
author={Qi Wei and Shuo He and Enneng Yang and Tingcong Liu and Haobo Wang and Lei Feng and Bo An},
booktitle={Forty-second International Conference on Machine Learning},
year={2025},
url={https://openreview.net/forum?id=a02CH43z1G}
}

@InProceedings{Zhu_2021_DFKD,
  title = 	 {Data-Free Knowledge Distillation for Heterogeneous Federated Learning},
  author =       {Zhu, Zhuangdi and Hong, Junyuan and Zhou, Jiayu},
  booktitle = 	 {Proceedings of the 38th International Conference on Machine Learning},
  pages = 	 {12878--12889},
  year = 	 {2021},
  editor = 	 {Meila, Marina and Zhang, Tong},
  volume = 	 {139},
  series = 	 {Proceedings of Machine Learning Research},
  month = 	 {18--24 Jul},
  publisher =    {PMLR},
  url = 	 {https://proceedings.mlr.press/v139/zhu21b.html}
}

@InProceedings{He_2016_ResNet,
author = {He, Kaiming and Zhang, Xiangyu and Ren, Shaoqing and Sun, Jian},
title = {Deep Residual Learning for Image Recognition},
booktitle = {Proceedings of the IEEE Conference on Computer Vision and Pattern Recognition (CVPR)},
month = {June},
year = {2016}
}

@inproceedings{Devlin_2019_BERT,
    title = "{BERT}: Pre-training of Deep Bidirectional Transformers for Language Understanding",
    author = "Devlin, Jacob  and
      Chang, Ming-Wei  and
      Lee, Kenton  and
      Toutanova, Kristina",
    editor = "Burstein, Jill  and
      Doran, Christy  and
      Solorio, Thamar",
    booktitle = "Proceedings of the 2019 Conference of the North {A}merican Chapter of the Association for Computational Linguistics: Human Language Technologies, Volume 1 (Long and Short Papers)",
    month = jun,
    year = "2019",
    address = "Minneapolis, Minnesota",
    publisher = "Association for Computational Linguistics",
    url = "https://aclanthology.org/N19-1423/",
    doi = "10.18653/v1/N19-1423",
    pages = "4171--4186"
}

@InProceedings{Yin_2020_DeepInversion,
author = {Yin, Hongxu and Molchanov, Pavlo and Alvarez, Jose M. and Li, Zhizhong and Mallya, Arun and Hoiem, Derek and Jha, Niraj K. and Kautz, Jan},
title = {Dreaming to Distill: Data-Free Knowledge Transfer via DeepInversion},
booktitle = {Proceedings of the IEEE/CVF Conference on Computer Vision and Pattern Recognition (CVPR)},
month = {June},
year = {2020}
}

@techreport{Krizhevsky_2009_CIFAR,
  title={Learning multiple layers of features from tiny images},
  author={Krizhevsky, Alex and Hinton, Geoffrey and others},
  year={2009},
  institution={University of Toronto}
}

@article{Alam_2018_CrisisMMD, title={CrisisMMD: Multimodal Twitter Datasets from Natural Disasters}, volume={12}, url={https://ojs.aaai.org/index.php/ICWSM/article/view/14983}, DOI={10.1609/icwsm.v12i1.14983}, abstractNote={ &lt;p&gt; During natural and man-made disasters, people use social media platforms such as Twitter to post textual and multimedia content to report updates about injured or dead people, infrastructure damage, missing or found people, among other information types. Studies have revealed that this online information, if processed timely and effectively, is extremely useful for humanitarian organizations to gain situational awareness and plan relief operations. In addition to the analysis of textual content, recent studies have shown that imagery content on social media can boost disaster response significantly. Despite extensive research that mainly focuses on textual content to extract useful information, limited work has focused on the use of imagery content or the combination of both content types. One of the reasons is the lack of labeled imagery data in this domain. Therefore, in this paper, we aim to tackle this limitation by releasing a large multi-modal dataset from natural disasters collected from Twitter. We provide three types of annotations, which are useful to address a number of crisis response and management tasks for different humanitarian organizations. &lt;/p&gt; }, number={1}, journal={Proceedings of the International AAAI Conference on Web and Social Media}, author={Alam, Firoj and Ofli, Ferda and Imran, Muhammad}, year={2018}, month={Jun.} }

@article{Li_2021_FedBN,
  title={Fedbn: Federated learning on non-iid features via local batch normalization},
  author={Li, Xiaoxiao and Jiang, Meirui and Zhang, Xiaofei and Kamp, Michael and Dou, Qi},
  journal={arXiv preprint arXiv:2102.07623},
  year={2021}
}

@inproceedings{Li_2020_FedProx,
 author = {Li, Tian and Sahu, Anit Kumar and Zaheer, Manzil and Sanjabi, Maziar and Talwalkar, Ameet and Smith, Virginia},
 booktitle = {Proceedings of Machine Learning and Systems},
 editor = {I. Dhillon and D. Papailiopoulos and V. Sze},
 pages = {429--450},
 title = {Federated Optimization in Heterogeneous Networks},
 url = {https://proceedings.mlsys.org/paper_files/paper/2020/file/1f5fe83998a09396ebe6477d9475ba0c-Paper.pdf},
 volume = {2},
 year = {2020}
}

@article{Sarker_2021_ML,
  author    = {Iqbal H. Sarker},
  title     = {Machine Learning: Algorithms, Real-World Applications and Research Directions},
  journal   = {SN Computer Science},
  volume    = {2},
  number    = {3},
  pages     = {160},
  year      = {2021},
  month     = mar,
  doi       = {10.1007/s42979-021-00592-x},
  url       = {https://doi.org/10.1007/s42979-021-00592-x},
  issn      = {2661-8907}
}

@article{Yang_2024_Model,
  title={Model merging in llms, mllms, and beyond: Methods, theories, applications and opportunities},
  author={Yang, Enneng and Shen, Li and Guo, Guibing and Wang, Xingwei and Cao, Xiaochun and Zhang, Jie and Tao, Dacheng},
  journal={arXiv preprint arXiv:2408.07666},
  year={2024}
}

@ARTICLE{Liu_2020_Hete,
  author={Liu, Feng and Zhang, Guangquan and Lu, Jie},
  journal={IEEE Transactions on Neural Networks and Learning Systems}, 
  title={Heterogeneous Domain Adaptation: An Unsupervised Approach}, 
  year={2020},
  volume={31},
  number={12},
  pages={5588-5602},
  doi={10.1109/TNNLS.2020.2973293}}
}

\end{document}